\documentclass[11pt]{article}
\usepackage{moriond,epsfig,graphicx}

\newcommand {\um}         {\, \mu \rm m}
\newcommand {\cm}         {\rm \, cm}
\def \e             {{\mathrm e}}
\def \Z             {{\mathrm Z}}
\newcommand {\ee}    {{\e ^+ \e ^-}}
\newcommand {\gtrsim}
     {\,\raisebox{-0.6ex}{$\stackrel{\textstyle>}{\textstyle\sim}$}\,}
\newcommand {\nm}         {\rm \, nm}
\newcommand {\ns}         {\rm \, ns}
\begin{document}
\vspace*{4cm}
\title{AN UNEXPECTED EFFECT IN ALEPH: LONG-TERM DISPLACEMENT
  OF THE SILICON VERTEX DETECTOR}

\author{ Giacomo Sguazzoni }

\address{European Laboratory for Particle Physics (CERN), CH--1211
  Geneva 23, Switzerland}

\maketitle\abstracts{
  The ALEPH Silicon Vertex Detector for LEP2 featured a
  laser survey system to monitor its mechanical stability. The analysis of 
  laser system data from 1997 to 2000 showed that VDET suffered a
  time-dependent displacement. It resulted to be compatible with a
  deformation of the support structure that made the device to slowly
  rotate during the data-taking. A~maximal local displacement of
  $\sim$20$\um$ was observed, corresponding to a rotation of
  $\sim$$10^{-4}{\rm rad}$.~The implementation of a time-dependent
  correction on the alignment by using the laser system data led to
  sizeable improvements on the ALEPH data quality.} 

\section{The ALEPH Silicon Vertex Detector}

Many physics studies of the high energy phase of LEP (LEP2) relied
on powerful heavy-flavour tags provided by high-resolution
vertex  detectors.

The ALEPH Silicon Vertex Detector$\,$\cite{vdetref} (VDET) had
an active length of $\sim$40$\cm$ and consisted of two concentric cylindrical
layers of 144 micro-strip silicon detectors of $\sim$5.3$\cm
\times$6.5$\cm$ with double-sided readout. The cylinders' axes
coincided with the nominal $\ee$ beam axes, i.e.~the $z$-axis in the
ALEPH reference system. Six silicon detectors were glued together and 
instrumented with readout electronics to form the VDET
elementary unit ({\em face}). The inner layer ($\sim$6.3$\cm$ radius)
was formed by 9 faces, the outer layer ($\sim$10.5$\cm$ radius)
consisted of 15 faces. 

\section{Laser System}

VDET featured a laser survey system$\,$\cite{laser1} to monitor its mechanical
stability with respect to the external tracking devices, since an
undetected large movement ($\gtrsim$20$\um$) of VDET during the
data-taking could have degraded significantly its performances. 
The VDET alignment was normally performed at the beginning
of the annual data-taking with events collected during
a dedicated calibration run at the Z resonance peak. Afterwards the reduced  
event rate of LEP2 energies did not allow a precise survey to be done by
using particle tracks, and the laser system thus played a crucial
role in monitoring the stability of the device.

\begin{figure}[t]
\begin{center}
    \leavevmode
    \begin{picture}(0,0)
      \put(18,20){\mbox{\small (a)}}
    \end{picture}
    \epsfig{file=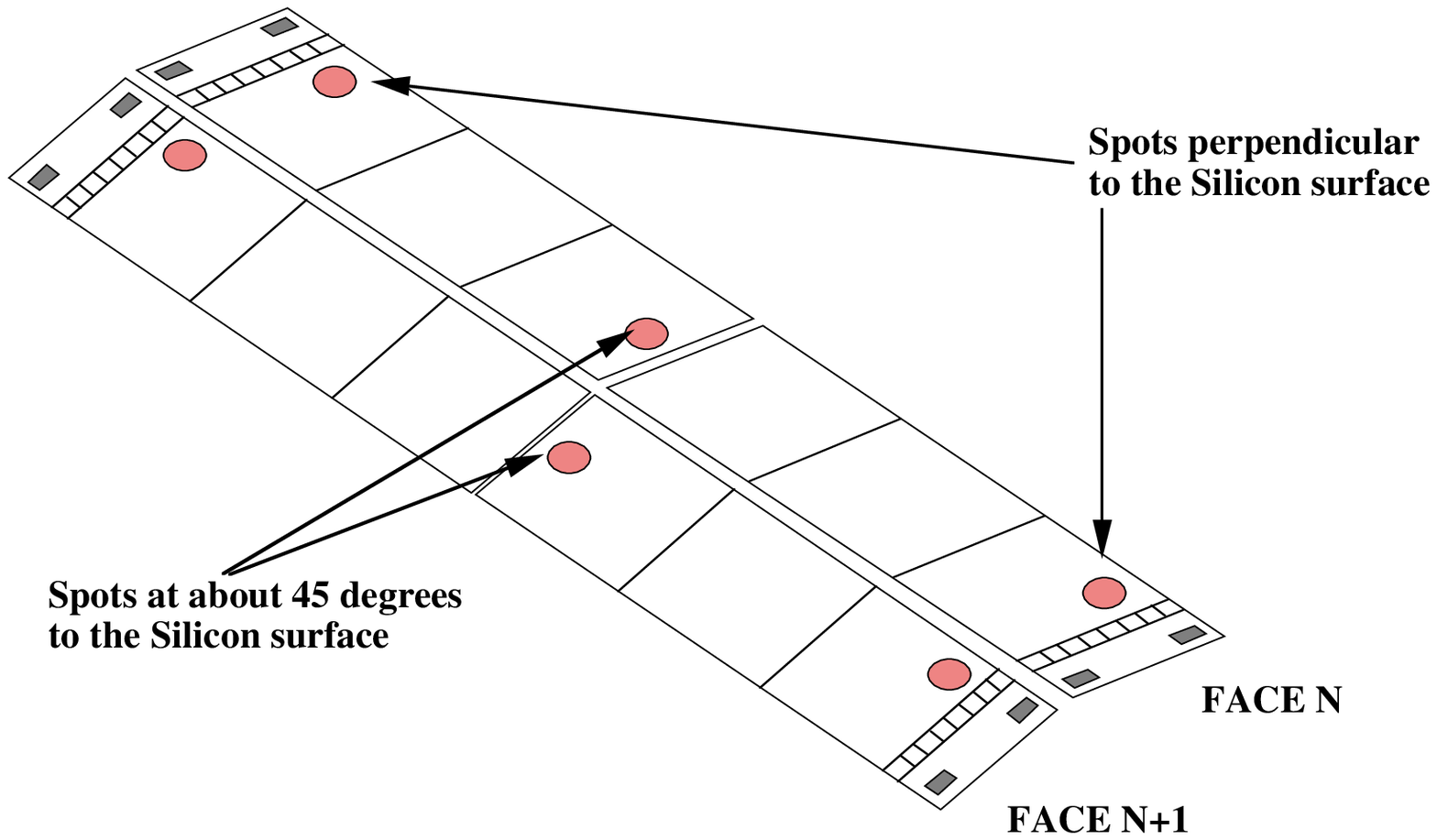, width=0.60\textwidth}
    \begin{picture}(0,0)
      \put(23,20){\mbox{\small (b)}}
    \end{picture}
    \epsfig{file=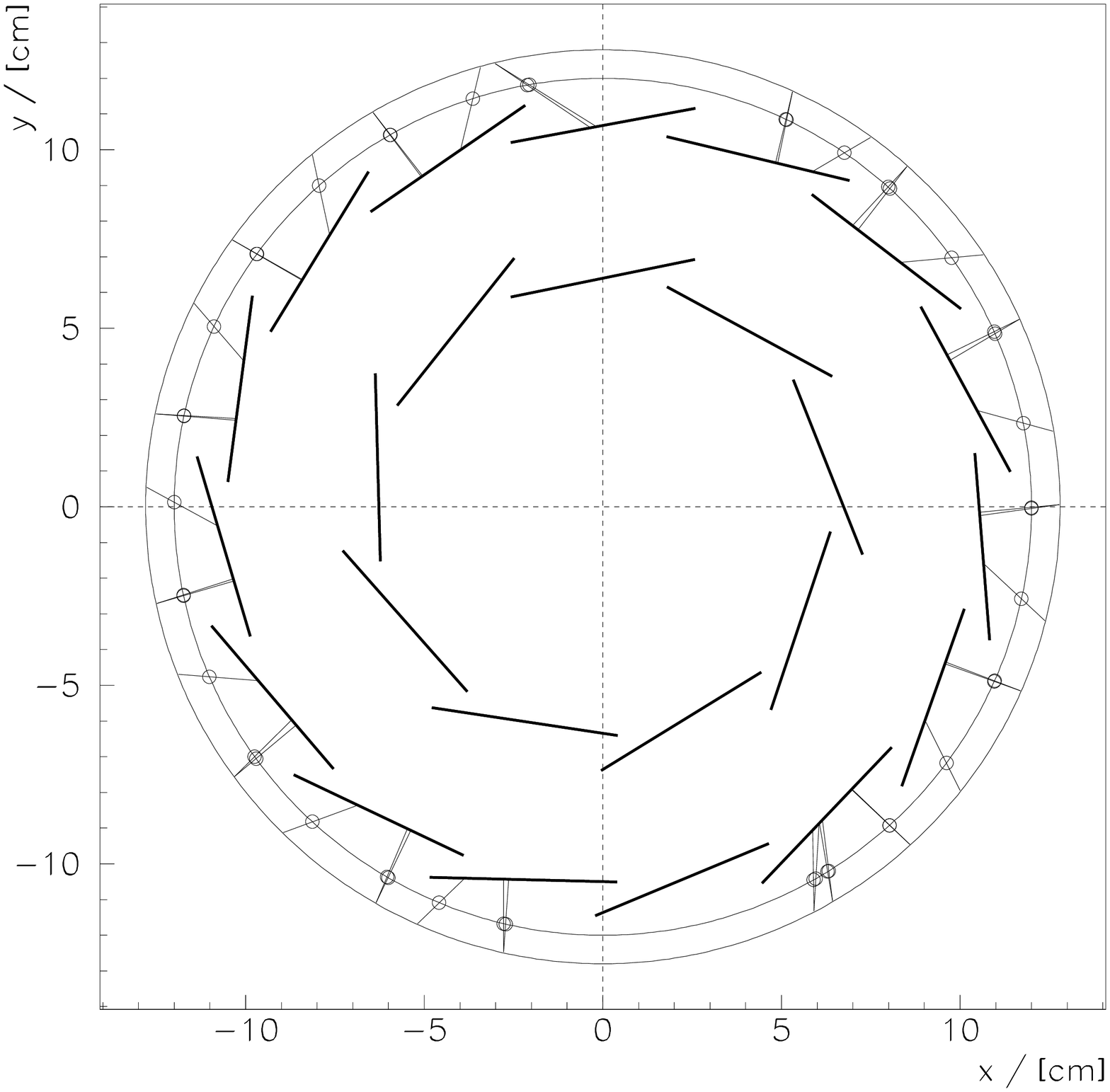, width=0.37\textwidth}
    \caption{(a) Sketch of laser spots on two VDET faces; (b) VDET
    ($r\phi$)-section with laser beams.} 
    \label{fig:vdetface}
    \label{fig:vdetsec}
\end{center}
\end{figure}

The system made use of infrared light ($\lambda = 904\nm$)
from two pulsed laser diodes with an output power of $6\,$W
each and a pulse length of $50\ns$.
The light was distributed via optical fibers to prisms and lenses 
attached to the inner wall of the Inner Tracking Chamber (ITC, the closest
outer detector). The lenses focused several light beams
on 14 of the 15 VDET outer faces. Information on the VDET
displacements with respect to the ITC were obtained by monitoring the
laser beam impact position ({\em spot}\/) on the silicon wafers.
All laser beams were nominally parallel
to the ($xy$)-plane. As shown in Fig.~\ref{fig:vdetface}a
and in Fig.~\ref{fig:vdetface}b, there were normally three spots per face: the
ones close to the face ends had the beam direction normal to the silicon
surface; the one placed about in the middle had the beam direction
at $\sim$45$^\circ$ with respect to the silicon surface, in order to be 
also sensitive to displacements normal to the face.

The system was operating during standard data-taking: the lasers were
fired approximately every $\sim$100 physics triggers (i.e.~once per one
to two minutes) and the spots were reconstructed as standard particle
hits. The spot position resolution was impressive, typically
$\sim$0.5--$1.5 \um$, thanks to the large signal and the cluster width
extending over more than one readout strip. After the installation five
spots out of 44 were missed, probably due to misalignment or breakage
of the optics. The number of collected laser events was 62k in 1997,
129k in 1998, 125k in 1999, and 170k in 2000.

\section{Analysis of Laser System Data}

\begin{figure}[t]
  \begin{center}
    \begin{picture}(0,0)
      \put(12,12){\mbox{\small (a)}}
      \put(146,12){\mbox{\small (b)}}
      \put(283,8){\mbox{\small (c)}}
    \end{picture}
    \epsfig{file=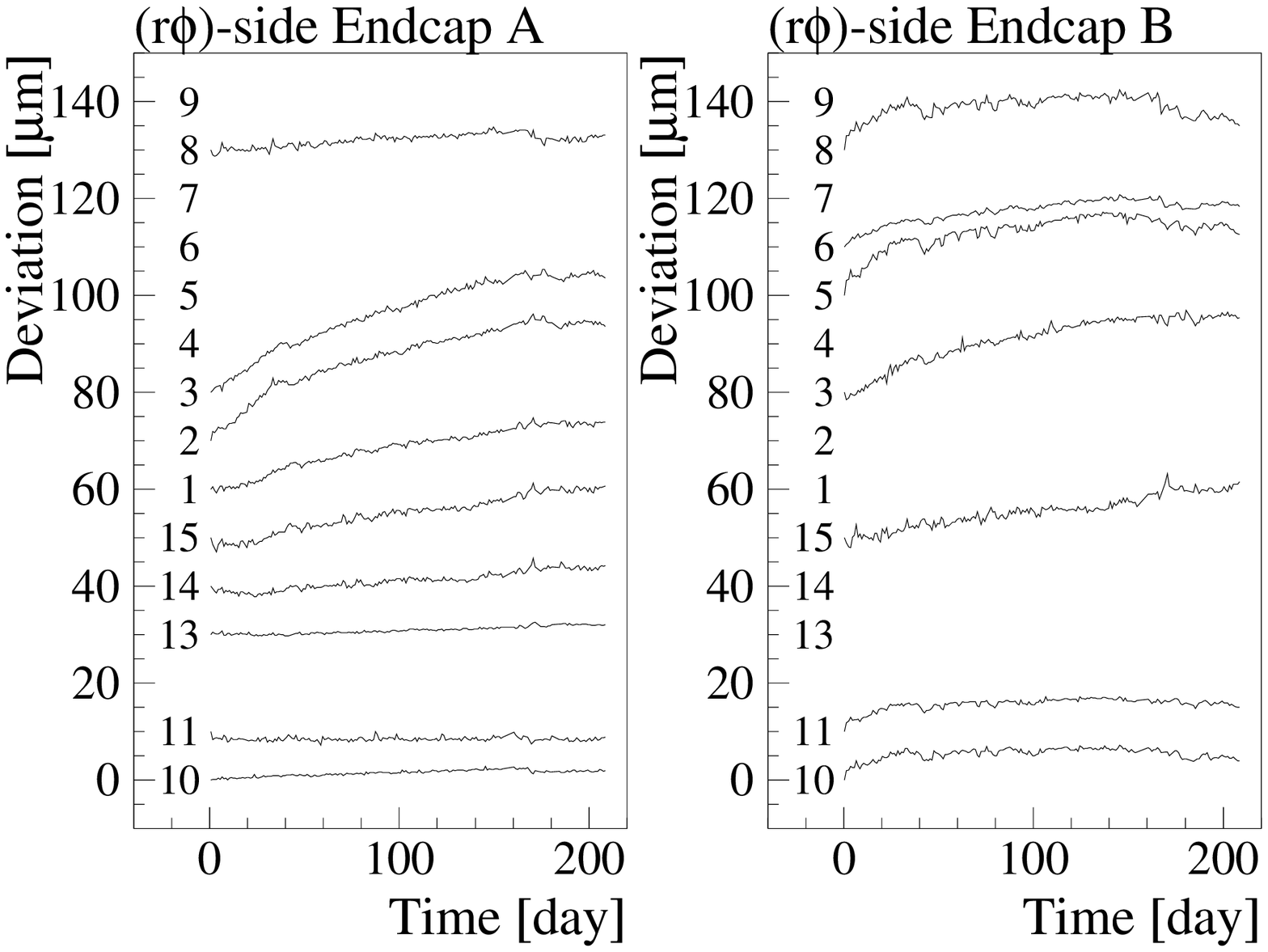, width=0.58\textwidth} 
    \hskip 3mm
    \includegraphics[bb=24 200 521 764,width=0.38\textwidth]{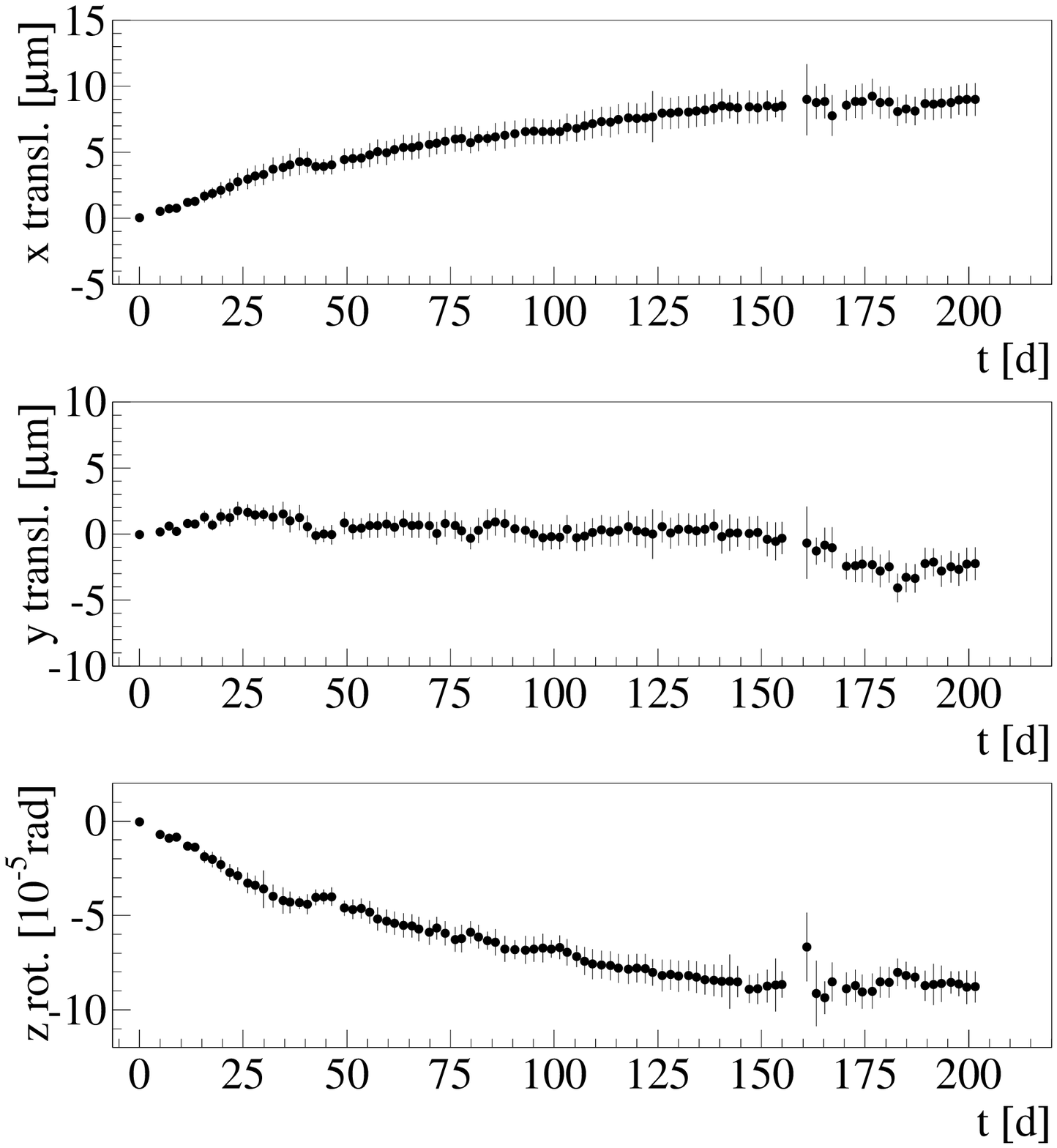}
    \caption{Time-charts of  ($r\phi$)-deviations for (a) endcap A ($z<0$)
    and (b) endcap B ($z>0$) spots in 2000 laser data. Each 
    time-chart is identified by the face number (as in
    Fig.~\ref{fig:reconstruction}) and shifted for graphical
    reasons. (c)~Two-days averaged correction parameters as
    a function of time from the beginning of the data-taking in 2000.}  
    \label{fig:rawtc} 
    \label{fig:parameters}
  \end{center}
\end{figure}

The analysis of the laser system data consisted in studying the time
dependence of the spot positions. Along with the expected short-term
displacements due to temperature effects$\,$\cite{laser2}, a long-term
dependence was discovered in ($r\phi$)-spot positions since 1997. As an
example, this behaviour over the entire data-taking
period in 2000 is shown in Fig.~\ref{fig:rawtc}a and in Fig.~\ref{fig:rawtc}b,
referring to endcap A ($z<0$) and endcap B ($z>0$) spots,
respectively. Some time-chart is not present because the corresponding
spot was missing or removed from the analysis because of low
quality$\,$\cite{laser2}. 
The size of the ($r\phi$)-displacements was systematically dependent on
the face azimuthal position, in some case being as large as
$\sim$10--20$\um$, thus comparable to the charged track single hit
resolution ($\sim$10$\um$). On the contrary, the $z$-side deviations did not
show any critical~effect. 

\begin{figure}[t]
  \begin{center}
    \hskip -2mm
    \begin{picture}(0,0)
      \put(20,20){\mbox{\small (a)}}
      \put(177,20){\mbox{\small (b)}}
      \put(337,86){\mbox{\small (c)}}
      \put(337,2){\mbox{\small (d)}}
    \end{picture}
    \includegraphics[bb=30 145 562 686, height=0.365\textwidth]{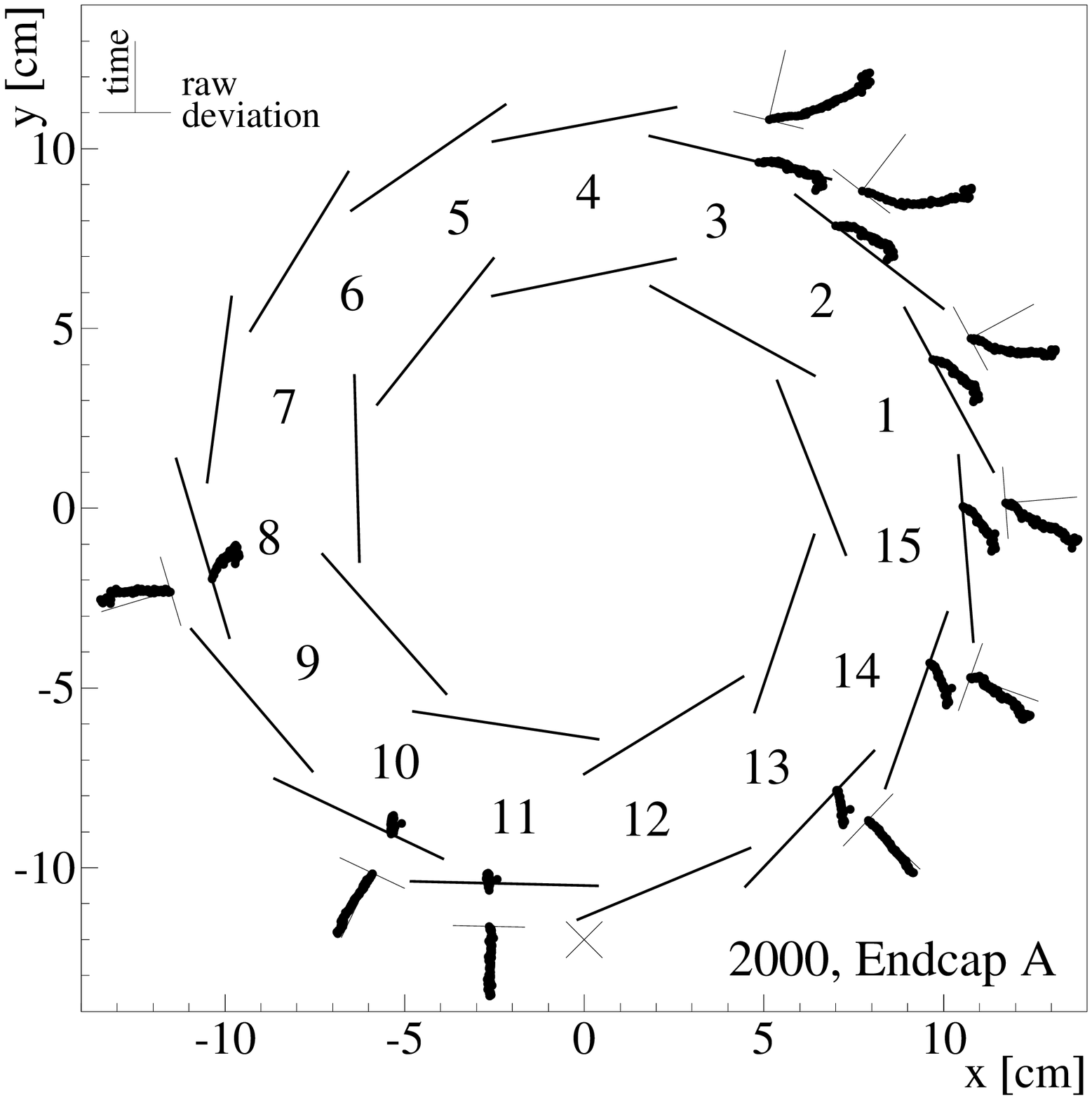}
    \includegraphics[bb=60 145 562 686, height=0.365\textwidth]{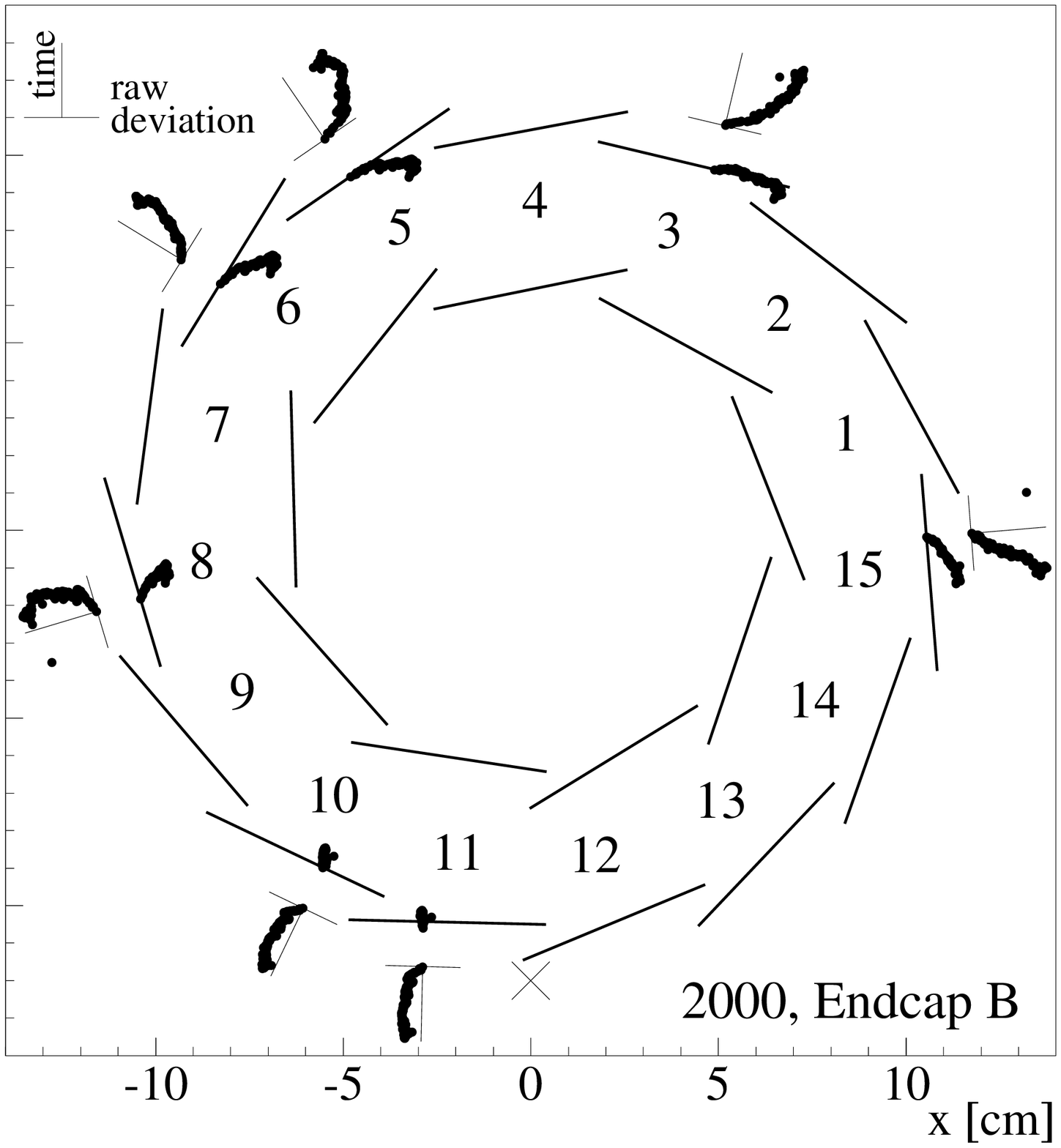}
    \hskip 2mm
    \epsfig{file=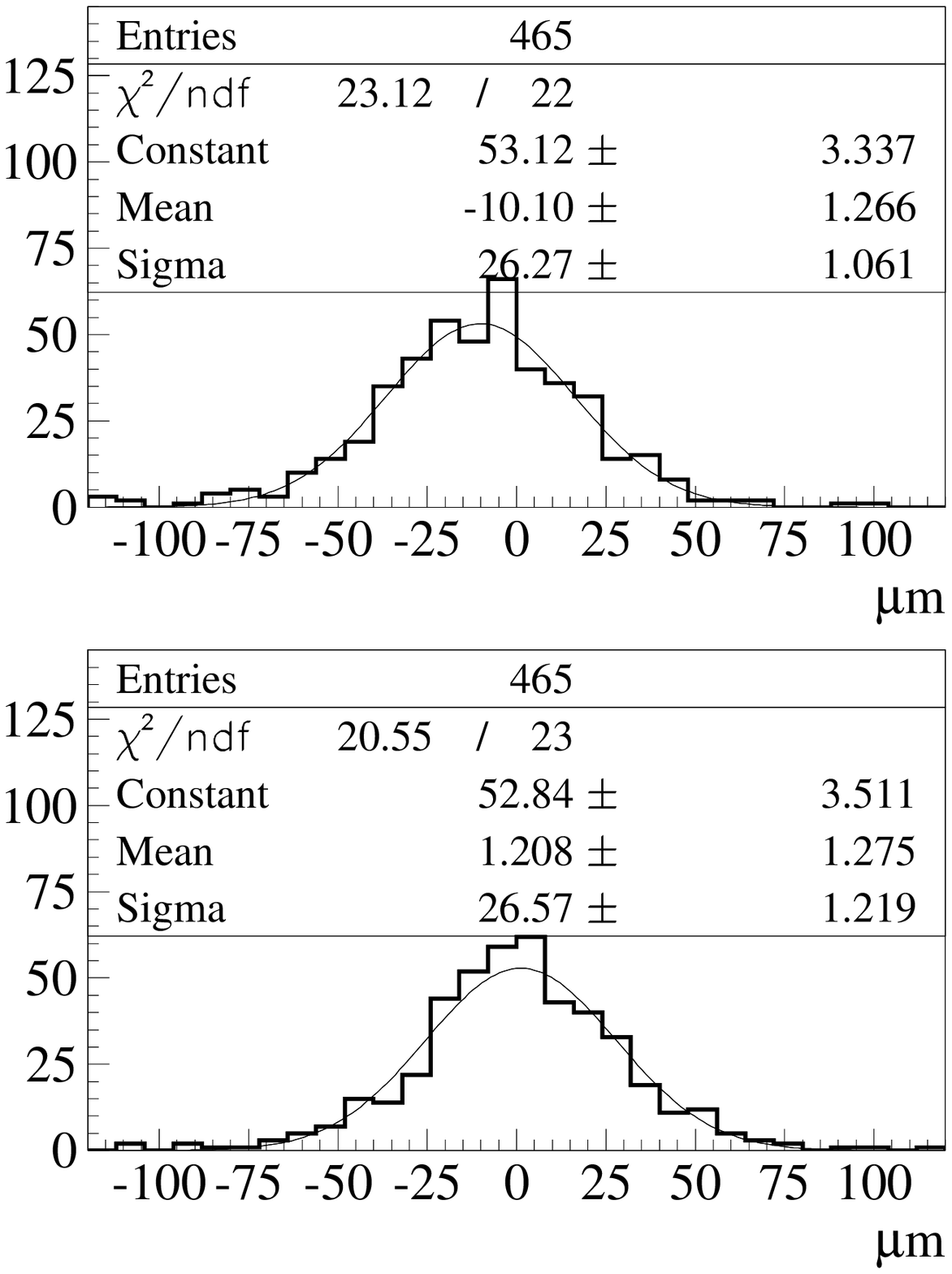, height=0.36\textwidth}
    \caption{Pictorial view of the reconstructed displacements
    for (a) endcap A and (b) endcap B spots: the dots represent their
    positions during time. The scatter plots outside the outer layer
    are the raw deviation time-charts of the corresponding spot. All
    displacements are amplified by a factor 1000. The cross is the
    probable rotation center. 
    (c) Distribution of sum of impact parameters of the two
    tracks in $\Z\to\mu\mu$ events before the
    laser correction to the alignment and (b) after the alignment correction.
    The plots refers to a Z calibration run taken in year 2000,
    approximately 150 days after the beginning of data-taking.} 
    \label{fig:reconstruction}
    \label{fig:sumd0}
  \end{center}
\end{figure}
%


The long-term deviations were likely to be due to a global rotation of
VDET in the ($xy$)-plane, as suggested by the distinctive geometrical
configuration. 
The deviations were thus parametrized under the assumption that
VDET was moving as a rigid body, i.e.~by using three parameters to correct
the nominal alignment: the $x$-translation, the $y$-translation and the rotation
angle around the $z$-axis. These were extracted by a fit
procedure. As an example, Fig.~\ref{fig:parameters}c reports the
parameters relative to year 2000 plotted as a function of the time
from the beginning of the data-taking. A pictorial view of the corresponding VDET
displacement is shown in Fig.~\ref{fig:reconstruction}a and in
Fig.~\ref{fig:reconstruction}b. The reconstructed motion was
consistent with a rotation and the rigid body model was found to
reproduce the observed spot deviations within a few 
microns$\,$\cite{laser2}. The maximum $z$-rotation value was
$\sim$7$\cdot10^{-5}$rad in 1997, $\sim$8$\cdot10^{-5}$rad in 1998,
$\sim$3$\cdot10^{-5}$rad in 1999 and $\sim$9$\cdot10^{-5}$rad in 2000.    

During the last three years of LEP running, several runs at the Z resonance peak
were also taken for calibration purpose, during and at the
end of the data-taking. The standard alignment performed by using
these calibration data allowed to independently confirm the motion of
VDET as extracted from the laser system, definitely proving its
reliability. A time-dependent correction of the alignment based on the
laser data was thus applied from 1998 on. 

The correction had a sensible
impact on the data quality. This was pointed out studying
the Z calibration runs, where a significant amount of data was 
collected in a period of time much shorter than the time-scale of the
effect under observation. Figure~\ref{fig:sumd0}c and Fig.~\ref{fig:sumd0}d
shows the distribution of the sum of the impact parameters of the two
muons in $\Z\to\mu\mu$ events, before and after applying the laser
correction. The plots refer to the Z calibration run taken in 2000,
$\sim$150~days after the beginning of data-taking, when the
displacement averaged over all VDET faces was about
$\sim$10$\um$. Without the correction the systematic shift of the   
distribution is of the same order of magnitude. Applying the laser
correction the mean of the distribution is again compatible with
zero.

The rotation may be explained considering that VDET is supported by
flanges which slot into two long metal rails located at the top
and bottom of the ITC inner cylinder. 
The top flanges were not metallic as the bottom ones, but made of
springy plastic to compensate for distance variations between
the rails along which VDET slided during the installation. 
The VDET environmental conditions during running were quite different
with respect to the shutdown period, both in terms of temperature and
humidity. They probably had some long-term effect on the material of
the plastic flanges, causing a tiny deformation that made VDET to
slowly rotate around the lower rail (represented by the small crosses in
Fig. \ref{fig:reconstruction}a and in Fig.~\ref{fig:reconstruction}b).

As a further indication of the environmental conditions as the possible cause,
the deformation was found to recover during the shutdown. In the
1999--2000 shutdown VDET was not removed for maintenance allowing its
position to be monitored. At the beginning of the data-taking in 2000
the position resulted again the same as at the beginning of the
data-taking in 1999. 

Extrapolating the LEP experience to the next generation of
experiments, the design issues of the Silicon Tracker devices 
for LHC turn out to be extremely challenging. The huge scale
involved forces the effects as the one here described to be carefully
taken into account. In fact, as an example, studies for the CMS Tracker
support structure have shown that a variation of environmental humidity
may lead to material deformation with very long
time-scale$\,$\cite{cms}.

\section{Conclusions}

The laser system of the ALEPH Vertex Detector for LEP2 was a simple but
extremely effective survey system. It allowed a reliable and high-precision
monitoring of the VDET position during the data-taking revealing a tiny
long-term rotation. An alignment correction based on the   
information from this system was successfully applied.



\end{document}